\documentstyle[pre,aps,epsf]{revtex}
\begin{document}
\draft

\twocolumn[\hsize\textwidth\columnwidth\hsize\csname@twocolumnfalse\endcsname

\title{Dynamics of driven interfaces near
isotropic percolation transition}

\author{M.-P. Kuittu$^{1}$, M. Haataja$^{2}$, 
N. Provatas$^{3}$, and T. Ala-Nissila$^{1,4,5,*}$}

\address{
$^1$Helsinki Institute of Physics, 
\\P.O. Box 9 (Siltavuorenpenger 20 C), 
FIN--00014 University of Helsinki, Helsinki, Finland
}

\address{
$^2$Centre for the Physics of Materials,
Department of Physics, McGill University,
Rutherford Building, 3600 ru\'e University,
Montr\'eal, Qu\'eb\'ec H3A 2T8, Canada
}

\address{
$^3$Department of Physics and Mechanical Engineering,
University of Illinois at Urbana-Champaign,
Loomis Laboratory of Physics,
1110 West Green Street, Urbana, IL, 61801--3080}

\address{
$^4$Permanent address: 
Laboratory of Physics, Helsinki University of Technology,
P.O. Box 1100, FIN--02015 HUT, Espoo, Finland
}  

\address{
$^5$Department of Physics, Brown University, Box 1843, 
Providence, R.I. 02912--1843
}

\date{March 9, 1998}

\maketitle

\begin{abstract}

We consider the dynamics and kinetic roughening of interfaces
embedded in uniformly random media near percolation treshold. 
In particular, we study simple 
discrete ``forest fire'' lattice models   
through Monte Carlo simulations 
in two and three spatial dimensions. An interface 
generated in the models
is found to display complex behavior. 
Away from the percolation transition, the interface 
is self-affine with asymptotic dynamics consistent with the  
Kardar-Parisi-Zhang universality 
class. However, in the vicinity of the percolation transition, 
there is a different behavior at earlier times.
By scaling arguments we show that the global scaling 
exponents associated
with the kinetic roughening of the interface can be
obtained from the properties of the underlying
percolation cluster. Our numerical results are in
good agreement with theory. However, 
we demonstrate that at the depinning transition, the 
interface as defined in the models is no longer self-affine. 
Finally, we compare these  
results to those obtained from 
a more realistic reaction-diffusion model of slow
combustion.

\end{abstract}


\vskip1pc]
\narrowtext

\section{Introduction}

Interfaces embedded in random media have received
a considerable amount of interest recently. Such 
diverse phenomena as pinning of flux lines in superconductors, 
dynamics of flame fronts in paper
and imbibition all contain interfaces 
propagating in random media with quenched noise
\cite{barabasi95}. For many such cases, an equation of motion
for the $d$ dimensional height variable $h(\vec r,t)$
can be written in the form

\begin{equation}
\frac{\partial h(\vec r,t)}{\partial t} =
\nu \nabla^2 h(\vec r,t) + \frac{1}{2} \lambda
\vert \nabla h(\vec r,t) \vert^2
+ F + \eta(\vec r, h),
\end{equation}

where $F$ is the driving force, and the noise term $\eta$
represents quenched disorder and is sufficiently short-ranged.

The behavior of driven
interfaces near the depinning transition
$F \rightarrow F_c$
at which the interface ceases to propagate
and its average velocity $v$ approaches zero
has turned out to be nontrivial.
In particular, there are two important
universality classes
that many different models of interface
dynamics fall into at the
depinning transition, namely the
isotropic depinning (ID) or the directed
percolation depinning (DPD) cases
\cite{barabasi95,tang92,tang95}. Roughly
speaking, models whose microscopic
dynamics is isotropic belong to the 
ID universality class, and those 
with spatial anisotropy to the 
anisotropic universality classes, of which perhaps
the most common one is the DPD case \cite{tang95}.
These universality classes can be distinguished by
the values of the scaling exponents associated with
the interface near the transition, as well as the
behavior of the nonlinear term $\lambda$ in the 
equation of motion for the interface. For the ID case,
$\lambda$ is kinetically generated ($\lambda \sim v$)
and vanishes at the transition, while for the DPD
case this is no longer true.

In this work we report the results of extensive
numerical simulations of some simple  
``forest fire'' lattice models \cite{albinet86}
where an interface propagates 
in a uniformly random background of reactants with an average
concentration $0< c < 1$.
This is an interesting special case of
a motion of an interface through a background medium of
quenched noise, with the additional feature that there
is an underlying {\it isotropic percolation transition} 
at some finite density $c^*$. Below $c^*$, the interface
becomes pinned due to the percolation transition,
and one may expect novel features to arise
in this class of problems, which we call 
here {\it isotropic percolation depinning}. 
There is little work on the
dynamics of interfaces in such 
isotropic lattice models, in particular near percolation
\cite{barabasi95,amaral94}. These type of models are 
also interesting from the point of view of recent
theoretical \cite{provatas95a,provatas95b,karttunen97}
and experimental \cite{zhang92,maunuksela97}
studies of dynamics of slow combustion in random media. 
 
Our results indeed reveal interesting and complex
behavior in the dynamics of the interface. 
Above the depinning transition for $c > c^*$, 
the kinetic roughening of the interface is found to be
described asymptotically by the  
Kardar-Parisi-Zhang (KPZ) \cite{kardar86} 
universality class as described by Eq. (1) with
annealed, gaussian noise. 
Results consistent with the KPZ universality
class were also found in the simulations of Refs. 
\cite{provatas95a,provatas95b} 
of a more realistic continuum model of slow combustion. 
On approaching the percolation transition of the
underlying lattice, $\lambda$
seems to decrease, since the
nonlinear term is kinetically generated in the
present case. We find that in this regime, there is a
different early-time behavior. We show that in this case
the {\it global} scaling exponents 
characterising the kinetic roughening of the 
interface can be obtained 
by utilizing results of percolation theory. In particular, this
means that these exponents
are completely determined by the
properties of the percolation cluster, and the continuum description
of Eq. (1) must break down. Furthermore, we show 
that the interface at $c^*$ as defined in the models is 
no longer self-affine but seems to
show {\it multiscaling}, since roughness exponents as measured
numerically from different 
correlation functions differ \cite{leschhorn96}.

The results from the discrete model are
compared and contrasted to those obtained 
for a continuum phase-field model of slow combustion introduced
and studied in Refs. \cite{provatas95a,provatas95b,karttunen97}.
We find that at high concentrations well above $c^*$, the two
models display qualitatively similar
behavior. However, as $c \rightarrow c^*$, the
kinetic roughening of the interface is different in
the two models, in that there is no evidence of
crossover in the continuum model. 
We show through an analytic argument that
this is essentially due to the divergence of the width
of the front in the continuum model and can be understood
in the framework of mean field theory.

The remainder of this paper is
organized as follows.
In Section 2 the model is 
introduced and 
characterized in detail. In 
Section 3 the results
of extensive Monte Carlo simulations 
in two and three spatial dimensions
are presented.
Also, a theory to explain
the observed crossover in the 
dynamics is developed
in this Section. A comparison between 
the discrete and continuum models
is carried out. Finally, in Section 4 
we conclude and discuss our results.

\section{The ``Forest Fire'' Models}

We consider the following simple ``Forest Fire''
(FF) cellular automaton models \cite{albinet86} on square and simple
cubic lattices in two and three spatial dimensions, respectively.
The status of each lattice site can be one
of the following: (i) an empty site, 
(ii) a site occupied by an 
unburned tree, (iii) a site occupied by a burning 
tree, and (iv) a site occupied by a burned tree. 
Initially, a fraction $c$ ($0< c < 1$) of the sites are occupied
by a tree. The initial distribution of trees is
uniformly random with no spatial correlations.
In the 2D case
the lattice is of length $L$ in the $x$ direction
with periodic boundary conditions,
and $L'$ in the $y$ direction
with free boundary conditions.
In the 3D case the
lattice is of length $L$ also in the $z$ direction.
Unless otherwise stated, $L' \gg L$ \cite{note1}.

The front propagation is initiated at $t=0$ by igniting
all the trees at the bottom of the lattice ($y=0$ in 2D and
the $xz$ plane in 3D, respectively).
The dynamics of the model is defined by the following set of rules:
During one Monte Carlo time step (MCS), a burning tree ignites all the
unburned trees in a fixed, finite region around it
and becomes a burned tree.
In this work, we consider the nearest neighbor (NN) and 
next nearest neighbor (NNN) FF models in 2D, and the NN
model in 3D. A burned tree will remain as such and 
new trees will not be generated during simulations, 
in contrast to several versions of this basic
model that display Self-Organised Criticality (SOC) 
\cite{bak90}.
The position of the emerging interface $h(\vec{r},t)$ 
at column $\vec{r}$ is defined as 
the location of the highest burning
tree {\it or} the highest burned tree, if there 
are no burning trees in that column \cite{anotherrule}. 
We note that this definition is sufficient 
to make the interface single valued.

The continuum model for which we will also present some new
results has been introduced and studied in Refs. 
\cite{provatas95a,provatas95b,karttunen97}. Briefly,  
the model is based on a phase-field appraoch, utilizing 
a set of coupled PDE's describing
the evolution of a thermal diffusion field $T(x,y)$ coupled to 
a random reactants concentration field $c(x,y)$. The interplay 
between thermal dissipation and reaction-diffusion of 
heat generated by combustion determines the dynamics in the model.
To study front propagation, 
the set of equations is discretized on a 2D lattice
and solved numerically. In analogy with the FF model, 
the lattice sites are randomly filled
with reactants (``trees'') which 
``burn'' according to the kinetics defined by the PDE's.
The main difference with respect to CA type of models is
that not only is the dynamics more realistic, but that
the effective range of interactions is in part determined by local
combustion dynamics. Also, the interface in the continuum model is not
sharp, but can be defined through the local maximum of the temperature
field $T(x,y)$.

\section{Results}

In order to quantitatively characterize the kinetic
roughening 
of the interface, we have considered the following
quantities \cite{barabasi95,lopez97}. 
First, the {\it global} width $w(c,t,L)$ of the
interface is defined by

\begin{equation}
w^2(c,t,L) \equiv \langle \overline{[h(\vec{r},t) - 
\overline{h(\vec{r},t)}]^2} \rangle,
\end{equation}

where the overbar denotes a spatial average over the system
of size $L$, and brackets denote configuration averaging.
Correspondingly,  
the {\it local} width of the interface $w_{\ell}(c,t)$ can be
defined as

\begin{equation}
w_{\ell}^2(c,t) \equiv \langle \langle [h(\vec{r},t) - 
\langle{h(\vec{r},t)}\rangle_{\ell} ]^2 \rangle_{\ell} \rangle,
\end{equation}

where the notation $\langle \rangle_{\ell}$ 
now denotes spatial averaging over all subsystems
of size $\ell$ of a system of total size $L$.
For growing self-affine interfaces, both the global and local
widths satisfy the Family-Viscek scaling relation \cite{family85}
and have asymptotic behavior given by

\begin{equation}
w^2(t,L) \sim \left\{ \begin{array}{ll}
                  t^{2 \beta}, & \mbox{for $t \ll L^z$;}\\
            L^{2 \chi}, & \mbox{for $t \gg L^z$,}
            \end{array}
            \right.       
\end{equation}

and correspondingly for $w_{\ell}^2(t)$. The quantities $\beta$ and
$\chi$ define growth and roughness exponents, respectively, and
$\chi = \beta z$ \cite{anomalous}. We note that in addition
to using the width, scaling exponents can be obtained by using
the height-height difference correlation function

\begin{equation}
C(r,t) = \langle \overline{\delta h(\vec r_0,t_0) - \delta h(\vec r_0
+ \vec r, t_0 + t)]^2} \rangle,
\end{equation}

with $\delta h \equiv h - \bar h$, 
in the appropriate regimes \cite{barabasi95}.

\subsection{Dynamics of one dimensional interfaces}

In the FF models, the emerging ``fire'' sweeps through all the
sites that are connected by the nearest or next nearest
neighbor rule throughout
the system. The front motion can thus be sustained only
for lattices whose average concentration $c$ is at or beyond
the percolation treshold of a 2D square lattice, which is
known to be $c^* \approx 0.592746$ and $c^* \approx 0.407254$ for
the NN and NNN cases, respectively \cite{stauffer91}. 
Thus, the pinning
of the interface below $c^*$ is a direct consequence
of the static percolation transition, and we call
this phenomenom {\it isotropic percolation depinning}
(IPD) here. There are two competing length scales in the problem,
namely the correlation length associated with the
percolation transition $\xi(c)$, and
the lateral correlation length of the moving interface
$\xi_{||}(t)$ that grows like $t^{1/z}$ \cite{huom}.
In the vicinity of $c^*$

\begin{equation}
\xi(c) \sim (c-c^*)^{-\nu},
\end{equation}

where $\nu$ defines the (static)
correlation length exponent, and
$\nu = 4/3$ in 2D percolation \cite{stauffer91}.

In the regime where pinning effects can be neglected,
it has been demonstrated that the quenched noise in Eq. (1) 
crosses over to thermal noise \cite{barabasi95}.
In the FF models, this sitation is realized well above $c^*$,
where $\xi(c)$ is essentially of the order of the lattice constant
and $\xi(c) \ll \xi_{||}(t)$ readily holds. 
Indeed, in this regime we find that the
interface moves with a constant velocity, and
its global width roughens asymptotically
as given by Eq. (4), with
$\beta \approx 1/3$ and $\chi \approx 1/2$ in
accordance with the KPZ universality class. 
For example, for the NN model at $c=0.95$ we obtain 
$\beta = 0.33(1)$ for $L=20000$ and $\chi=0.50(2)$
for a system size of $L=5000$. 
In Fig. 1 we
plot $w(t)$ vs. $t$ and the effective exponent
$\beta_{\rm eff}(t)=\ln w(t) / \ln t$ that shows the
asymptotic KPZ behavior. We would like to point
out that there is an initial time regime where the width grows
according to the uncorrelated random deposition model,
with $w(t) \sim t^{1/2}$.   For the NN model, this regime
is long-lived only for $c$ very close to unity \cite{NNNref}.

The asymptotic KPZ behavior for $c > c^*$ is not unexpected,
since the velocity of the
interface in the FF models is clearly tilt-dependent,
which generates the nonlinear term proportional to
$\lambda$ in Eq. (1). In Fig. 2(a) we show the
behavior of $\lambda$
as a function of $c$ above $c^*$ for the NN case. It has been
calculated numerically by computing
the average velocity $v$ of the interface 
as function of
the global tilt $m$, and fitting
a parabole to it for $m \ll 1$ \cite{barabasi95,amaral94}.
The interesting result is that $\lambda$ displays
nonmonotonic behavior and seems to eventually
{\it decrease} when approching the percolation
transition. In fact, we expect that $\lambda(c)
\rightarrow 0$ as $c \rightarrow c^*$ because 
the interface is eventually forced to 
propagate in the infinite percolation cluster 
which is known to be self-similar 
and isotropic at the 
percolation threshold \cite{vicsek89}. 
Since there is no preferred 
growth direction at $c^*$, the tilting of 
the interface should not affect 
the velocity of the interface any more.

This diminishing of $\lambda$ on approaching
$c^*$ means that the nonlinear term in Eq. (1)
becomes less and less important at early times
where $\xi(c) \gg \xi_{||}$. For the continuum
description to hold, however, the local slopes of
the interface should also remain small. 
We have studied this numerically for
various values of $c$ close to $c^*$, where
two things can be observed for the behavior
of the global width $w(t,L)$. First, the
range of the late-time KPZ scaling regime becomes
smaller in time as $c^*$ is approached from above.
Second, another regime where well defined 
power-law scaling of $w(t) \sim t^{\beta^{*}}$ 
can be observed appears at earlier times.
In Fig. 3(a) we show the behavior of the global width
for a system of size $L=20000$ at $c=0.59275$ (NN model).
We find that starting from early times, there is
a scaling regime where the growth exponent
$\beta^* = 0.88(1)$. Simulations of the NNN model
at $c = 0.407$ give $\beta^* \approx 0.88$, correspondingly.
In this regime parts of
the interface become pinned by unburned regions
on the lattice, and the interface motion consists
of large jumps, with large local slopes appearing.
This behavior indicates that the interfaces
may not be self-affine \cite{leschhorn96}.

The numerically observed crossover behavior
can be formulated
theoretically by assuming that it is induced by
the underlying percolation transition.
We write the following scaling form for the 
global width $w(c,t)$:

\begin{equation}
w(c,t) = \xi(c) f(\frac{t}{\tau_c}),
\end{equation}

where $\tau_c$ denotes the crossover time 
to the KPZ regime, and
the scaling function $f(u)$ has the limits

\begin{equation}
f(u) \sim \left\{ \begin{array}{ll}
                    u^{\beta^*}, & \mbox{if $u \ll 1$;} \\
                    u^{\beta} , & \mbox{if $u \gg 1$.}
                    \end{array}
            \right.
\end{equation}

Here $\beta^* \approx 0.88$ and 
$\beta = 1/3$. Using Eq. (5) for 
$\xi(c)$ and assuming that 
$\tau_c(c) \sim (c-c^*)^{-\Delta}$ we find that
best data
collapse as shown in Fig. 4, is obtained for 
$\nu=1.3$ and $\Delta=1.65$. Taking $\nu=4/3$, and
the dynamic exponent $z^*$
given by the exponent $d_{\rm min}\approx 1.13$ 
associated with the scaling of the 
minimum path distance \cite{amaral94,havlin95},
$\Delta = \nu z^* \approx 1.51$. Thus, our numerical
results are in good agreement with theory.

At $c^*$ where $\xi_{||} \ll \xi(c)$ for long
times ($t \ll L^{z^*}$), 
the interface is pinned by 
clusters formed by
the unoccupied sites, and the quenched disorder dominates.
The interface follows the ``edge'' of the infinite percolation
cluster. The global roughness exponent $\chi^*$ can
be then be directly deduced from the geometric properties 
of the percolation transition. 
In particular \cite{amaral94,barabasi95},
$\chi^* = \nu_{\perp} / \nu_{||}$, where
$\nu_{\perp}$ and $\nu_{||}$ 
are the perpendicular and parallel correlation length 
exponents of the critical percolation cluster, 
respectively. Since the percolation cluster in
the FF model is isotropic and $\nu_{\perp} = \nu_{||} = \nu$,
the roughness exponent $\chi^* = 1$ {\it in all dimensions}.
In this case, the exponent $z^*=d_{\rm min} \approx 1.13$, which
leads to $\beta^*=\chi^*/z^* = 1/d_{\rm min} 
\approx 0.88$ in excellent agreement
with our simulations. These results indicate that the
continuum description of Eq. (1) must break down at $c^*$
for the present IPD case \cite{QEW}.
 
We have examined the interface roughness exponent
$\chi$ numerically by studying the interface dynamics
as close to $c^*$ as possible. 
We have computed the generalized $q$th order height
difference correlation functions

\begin{equation}
G^q(r,t) = \langle \overline{[h(r,t) - \overline{h}(t)]^q} \rangle
         \sim r^{q \chi_q}, \ {\rm for} \ r \ll \xi_{||},
\end{equation}

by running the simulation until the interface finally stops
(for a finite system)
and approximately traces out the edge of the percolation cluster.
For a self-affine interface there is only one roughness
exponent, and thus
$\chi=\chi_q$ for all $q=2,4,6,...$. Our numerical results for 
a $L=2000$ system at $c=0.5928$ (NN model) 
give that $\chi_2=0.54(5)$,
$\chi_4=0.29(3)$ and $\chi_6=0.21(2)$. This indicates that
the interface associated with the percolation cluster {\it as
defined in the model} is not self-affine at $c^*$.
The reason is most likely that
the overhangs in the front edge of the interface that
follow the percolation cluster, are removed.
However, the scaling exponents for each higher order
correlation function that we have calculated seem to
be very well defined, which is an indication of
multiscaling similar to that seen in the longitudinal structure
functions in the study of turbulence \cite{turbulence}.

We have also numerically 
verified the scaling 
of the average velocity $v(c)$ of the
interface as a function of $c-c^*$ 
near the percolation
threshold (see Fig. 5). It is expected
to vanish as 

\begin{equation}
v(c) = A (c-c^*)^{\theta}.
\end{equation}

Our data for the NN model
give $A\approx 1.14$ and our best
estimate for the velocity exponent is
$\theta = 0.169(5)$ (the NNN model gives
$\theta = 0.17(5)$). 
In order to check the consistency of
this result, we note that 
there exists
a well-known scaling relation 
between $\theta$,
$z^*$, $\nu$, and $\chi^*$, 
namely \cite{barabasi95}

\begin{equation}
\theta=(z^*-\chi^*)\nu.
\label{sca}
\end{equation} 

By using the values $z^* = 1.13$, 
$\chi^* =1 $, 
and $\nu = 4/3$ we obtain
$\theta = 0.173$. This is 
in very good agreement with our data.

\subsection{Dynamics of two dimensional interfaces}

The 3D lattice model that we have studied is a simple
generalization of the 2D case to a simple cubic geometry.
We only consider the NN case here. 
The behavior of the emerging surface near
the percolation threshold is 
qualitatively similar
to the 2D case. In particular, in the long
time limit for $c > c^*$ the interface 
roughens in time with the
growth exponent $\beta = 0.24(2)$ as shown
in Fig. 1(b), in excellent 
agreement with 
numerical solutions of 
the $d=2+1$ KPZ equation and various discrete
models that belong to the KPZ universality class
\cite{barabasi95,halpinhealy95,kpz2d}. 

Closer to $c^* \approx 0.316$, we see the percolation-induced 
crossover. At $c= 0.316$ we find numerically that 
the interface
roughens with a growth 
exponent $\beta^* = 0.72(5)$ (Fig. 3.(b)).
Again, the exponents characterising 
the interface can be
obtained from the exponents 
of the critical
percolation cluster. In particular, 
it is reasonable to assume that the global
roughness exponent $\chi^* = 1$ since the cluster
is isotropic. Moreover, the 
minimum path exponent is
known in $d=3$ to be $d_{\rm min}=1.38(2)$, 
and this determines the dynamic
exponent $z^*=d_{\rm min}$ \cite{havlin95}. As a
consistency check, taken
together with the result that $\chi^*=1$,
this implies that, at depinning transition,
$\beta^* = \chi^*/z^* \approx 0.724$. Our direct
evaluation of $\beta$ agrees very well with
this prediction. We expect again that the
interface as defined in the 3D model is not self-affine
at $c^*$; however, we have not 
computed $\chi^*$ numerically.

We have also calculated the velocity exponent
and find $\theta=0.26(2)$. Using the scaling exponent
relation $\theta = (z^*-\chi^*) \nu$ with
$z^*=1.38$, $\chi^*=1$, and $\nu=0.88$ gives
$\theta=0.33$ which is in reasonably
good agreement with our data.

\subsection{Flame front propagation in the continuum model}

We have compared the dynamics of interface of the lattice 
model to a more 
realistic continuum reaction-diffusion model of 
Refs. \cite{provatas95a,provatas95b}.
This model is a type of phase-field model that couples 
the evolution of a thermal diffusion field to a randomly
distributed concentration field of reactants. 
The model couples the effects of thermal 
dissipation and diffusion to 
heat generated by combustion, via an 
Arrhenius-activated reaction term.
To study front propagation, 
the model is discretized on a 2D lattice
and solved numerically. In analogy with the FF model, 
the lattice sites are randomly filled
with reactants (``trees''), with an average normalized
concentration of $c \equiv \overline{c(x,y)}$. After ignition
of the bottom row of reactants at $t=0$, the heat
generated will ignite other occupied lattice sites around it,
and the local field $c(x,y)$ corresponding to the sites of the ``burning''
reactants will quickly
approach zero as determined by the equations.
A single-valued interface in the model is defined by the maximum of
the temperature field $T(x,y)$ for each column $x$. 

Previously, it was shown that that the kinetic roughening 
of the flame fronts generated by the continuum model 
belong to the thermal KPZ universality class
\cite{provatas95a,provatas95b}. In the limit of
almost uniform background density, the KPZ description
was also derived analytically from the set of
equations for the model \cite{provatas95b}.
The main difference with respect to the FF lattice model
was that even very close to the percolation
treshold of the model $c^*\approx 0.20$, there was
no evidence of percolation induced crossover. 
Also, the
continuum model near $c^*$ gave results that
were consistent with the mean field theory of
percolation, {\it e.g.} $\nu \approx 0.5$ and
$\theta \approx 0.5$ \cite{provatas95b}.

In Fig. 2.(b) we show the behavior of the
nonlinear coefficient $\lambda$ for the continuum 
combustion model.
Similarly to the FF lattice model, we find that
$\lambda$ approaches zero for $c \rightarrow c^*$.
However, unlike the the lattice model, no crossover 
behaviour is observed as $c \rightarrow c^*$. This 
is explained as follows: From the mean-field 
analysis of Ref. \cite{provatas95b}, 
the leading front of the thermal field decays as

\begin{equation}
T_{MF}(x) \sim e^{-x/l_D},
\end{equation}

where $l_D=D/v_m$ is the thermal diffusion length 
defining the range of effective interactions in the model,
and thus also  
the scale of the intrinsic thickness of the interface 
$w_{\rm int}$. The constants 
$v_m$ and $D$ are the mean interface velocity and 
thermal diffusion constant, respectively.
Using the result that  $v_m \sim (c-c^*)^{0.5}$, we conclude
that $w_{\rm int} \sim (c-c^*)^{-0.5}$. On the other hand,
in the MF percolation transition
the correlation length scales as $\xi(c) \sim (c-c^*)^{-0.5}$.
These results imply that the thickness of the 
interface has {\it the same divergence} as the 
correlation length, within which the 
crossover behavior should be observed. 
Thus, everything 
that happens on length scales smaller 
than $w_{\rm int}$ will be smeared out.
Therefore, due to the
increasing thickness of 
the interface, the
second regime at early times is never observed.

\section{Summary and Discussion}

In this work we have studied the
dynamics of interfaces in random media
through Monte Carlo simulations of
some discrete cellular automaton models of
``forest fires''. We find
that away from the depinning transition 
induced by the isotropic percolation transition of
the underlying lattice, 
the kinetic roughening is asymptotically described by the
Kardar-Parisi-Zhang
\cite{kardar86}
universality class.
In the vicinity of the IPD transition, however, 
the behavior is found to be 
different. At the transition, 
the {\it global} roughness exponent
$\chi^*$ and the growth exponent $\beta^*$ 
are completely determined by the geometric properties
of the percolation transition, leading to the
result that $\chi^*=1$, $\beta^*=1/d_{\rm min}$ 
in all dimensions.
We have verified this numerically for the exponent
$\beta^*$ in 2D and 3D cases.
However, by computing the roughness exponent of the
interface from different correlation functions, we
find that the interface is no longer self-affine, but
seems to indicate multiscaling.
This is most likely due to the removal of overhangs
in the way the interface is defined in the models.

Comparison between the lattice 
models and the
more realistic model of Refs. 
\cite{provatas95a,provatas95b,karttunen97} 
was made, and qualitatively similar behavior
was found at high concentrations. 
Interestingly,
however, the two models
displayed qualitatively
different behavior for 
$c \rightarrow c^*$.
In particular, the exponents compatible with
the KPZ universality
were shown to hold for all 
values of $c$ studied
in Refs. \cite{provatas95a,provatas95b}. 
We demonstrate that this can be understood on a
basis on the mean-field 
nature of the percolation
transition exhibited by 
the continuum model. 

The models studied here are particularly interesting
from the point of view of the recent experiments on
slow combustion of paper \cite{maunuksela97,maunuksela98}.
In these experiments, asymptotic KPZ exponents were verified
for the first time for driven interfaces. This is in complete
agreement with all the models here well above percolation,
as well as the DPD universality class.
Near percolation, the assumption made on the basis of
the earlier experiments by Zhang {\it et al.} \cite{zhang92}
has been that DPD effects dominate \cite{barabasi95}.
However, the most recent experiments indicate \cite{maunuksela98}
that the effective short-range exponents before KPZ asymptotics
may not be well defined.

Acknowledgements: We wish to thank Mr. Tommi Karhela for
help at the initial stages of this work, and 
J. Krug for useful discussions.
This work has in part been supported by the Academy of Finland. \\



$^*$Corresponding author. E-mail address:
{\tt Tapio.Ala-Nissila@helsinki.fi}.

\pagebreak
\cleardoublepage
\pagebreak

\begin{center}
\Large 
{\sc Figure captions}
\end{center}

\normalsize

\vspace{2.0cm}

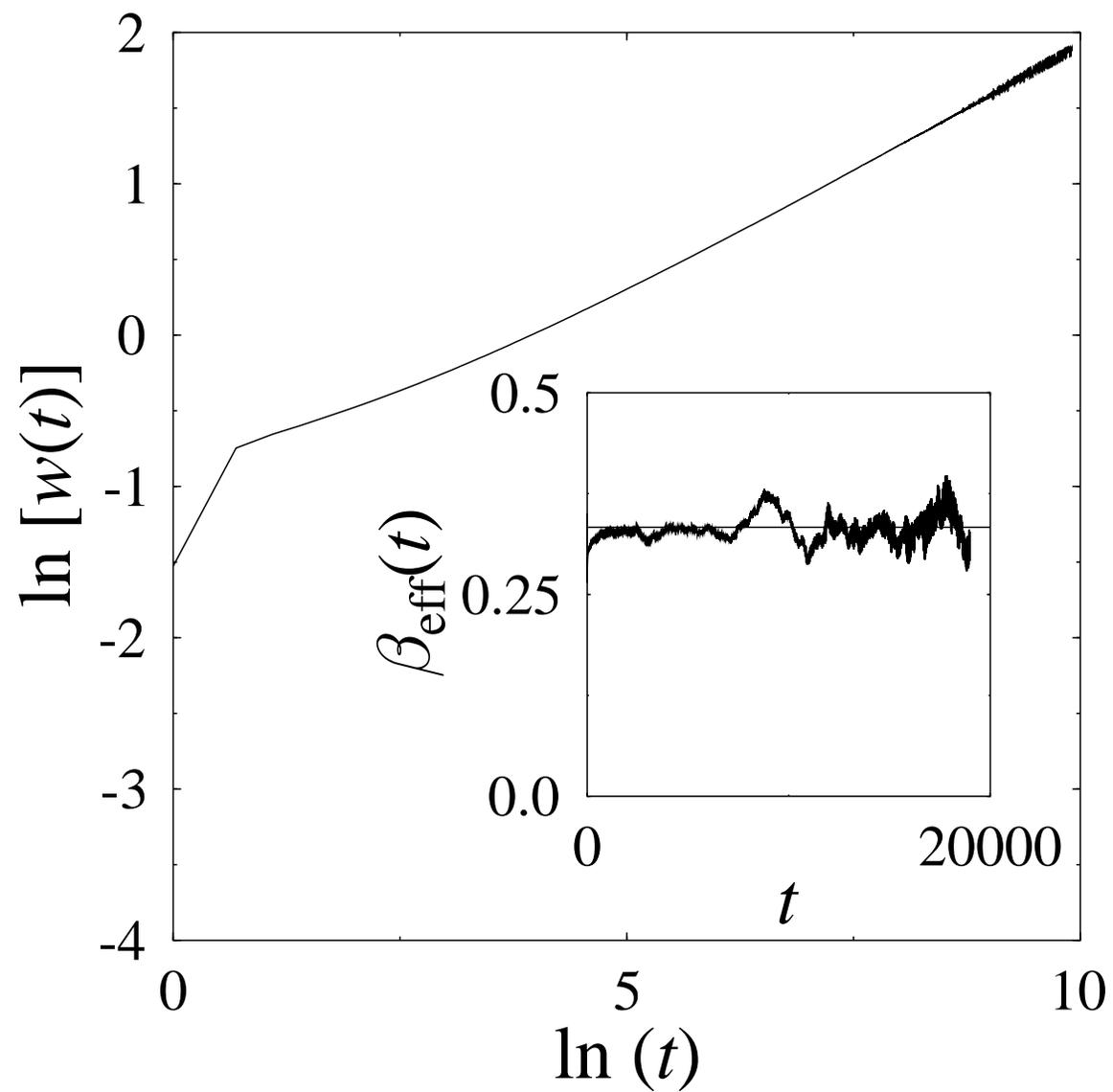
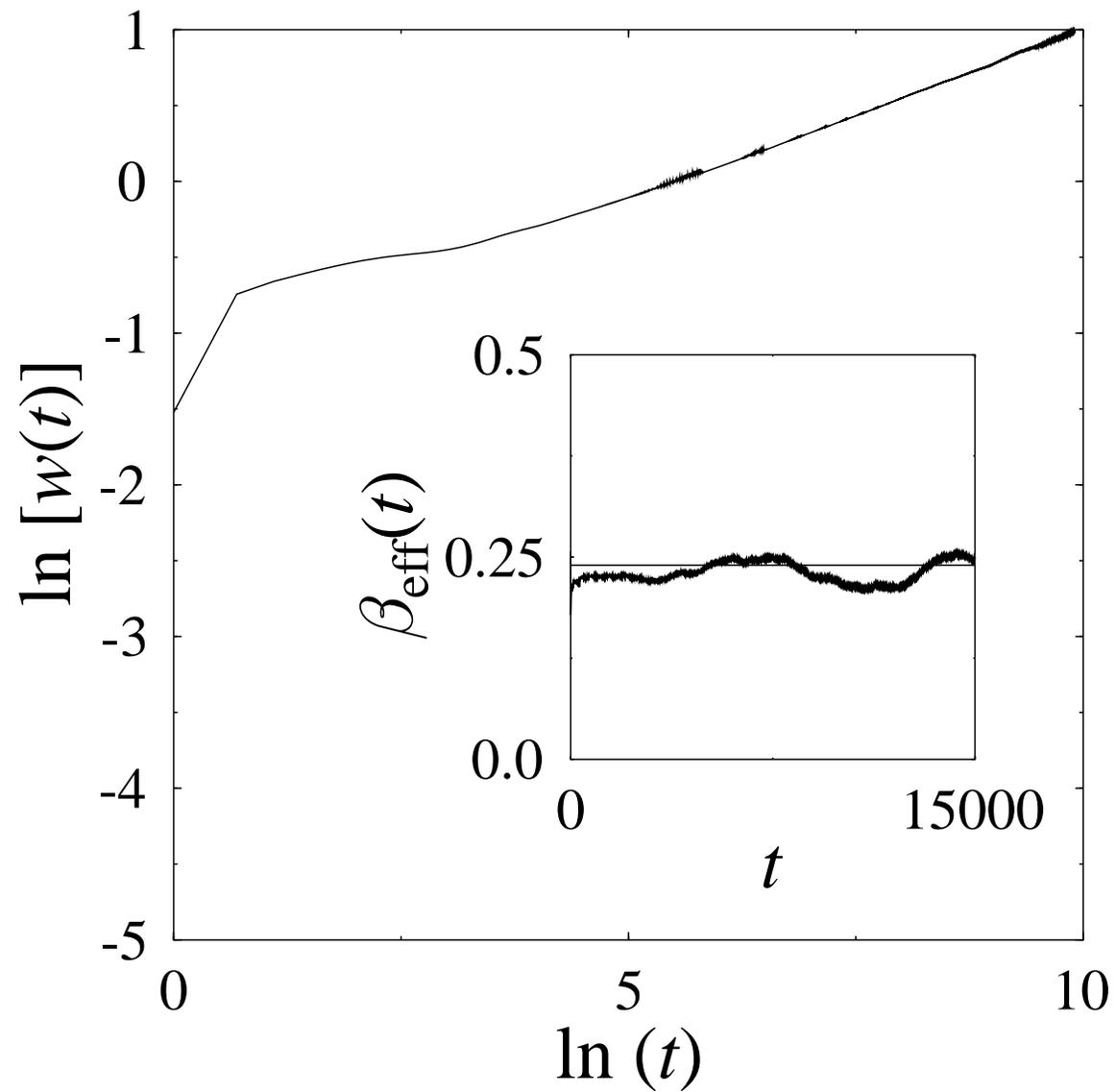
Fig. 1. (a) The global width $w(t)$ vs. $t$ 
in 2D for the NN model with
$c=0.95$ and $L=20000$. The inset shows the effective
growth exponent $\beta_{\rm eff}(t)$ vs. $t$, where
$\beta_{\rm eff}(t) \equiv \ln w(t) / \ln t$. The exact KPZ
value of $\beta=1/3$ is shown by the horizontal line.
(b) The global width $w(t)$ vs. $t$ for the 3D NN model, with
$c=0.97$ and $L \times L = 200 \times 200$. In both cases 
averages were taken over $100$ runs.
Inset shows $\beta_{\rm eff}(t)$, with
the horizontal value indicating the KPZ result $\beta \approx 
0.24$.

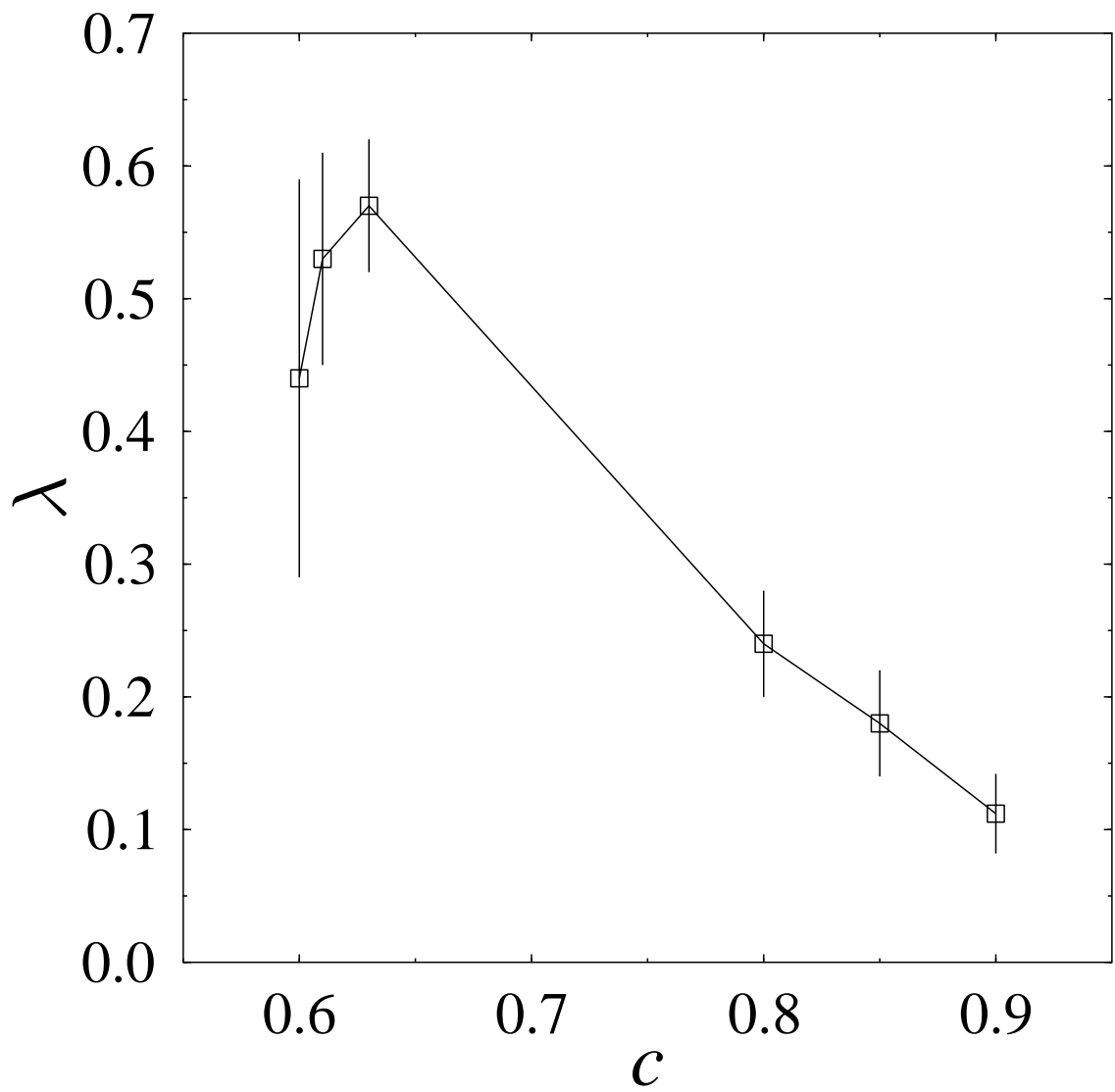
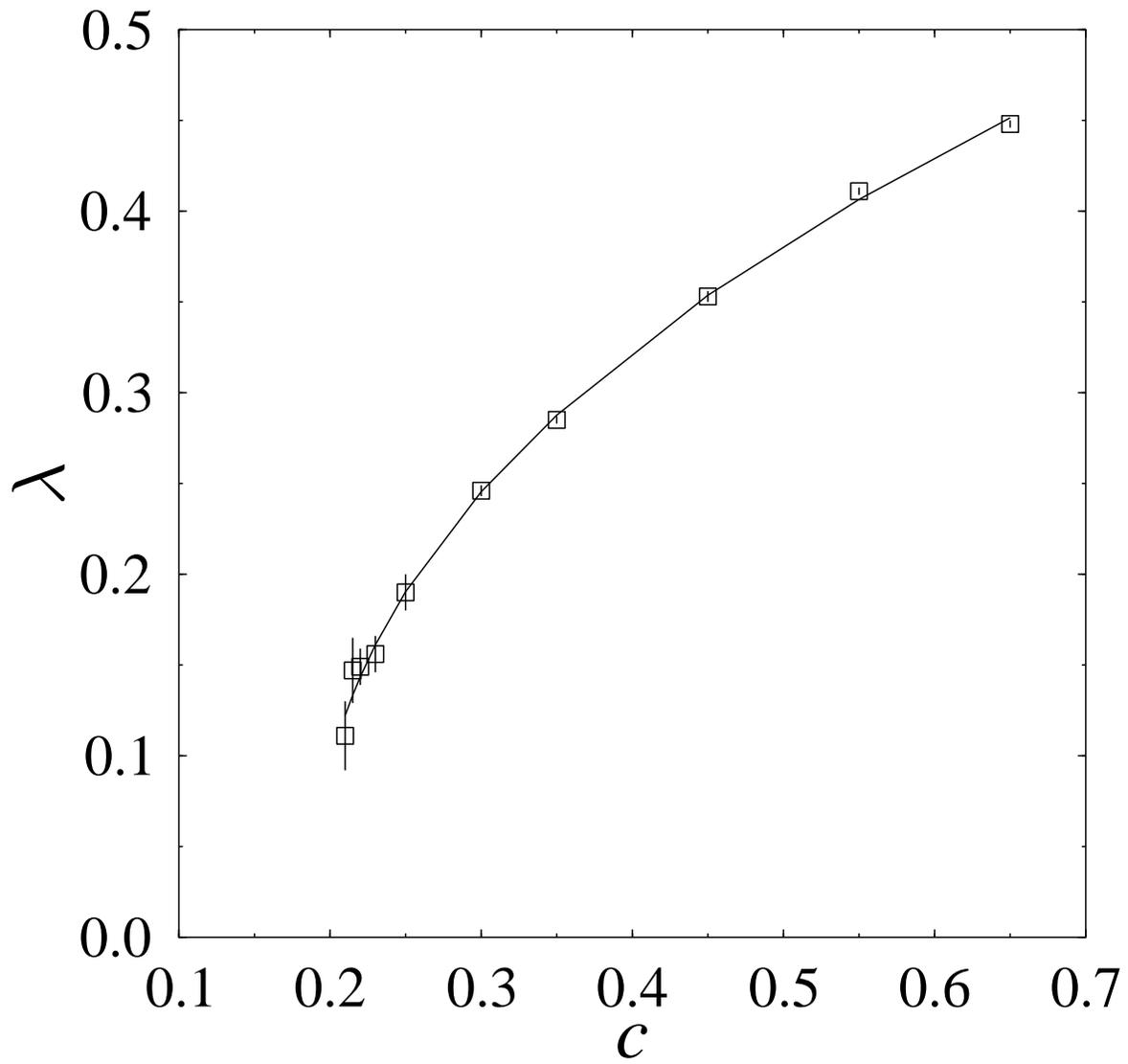
Fig. 2. (a) $\lambda$ vs. $c$ in 2D
with $L=2000$ for the NN lattice model.
The data was averaged over 1000 runs. (b) $\lambda$ vs. $c$
for the continuum model, with $L=200$. 

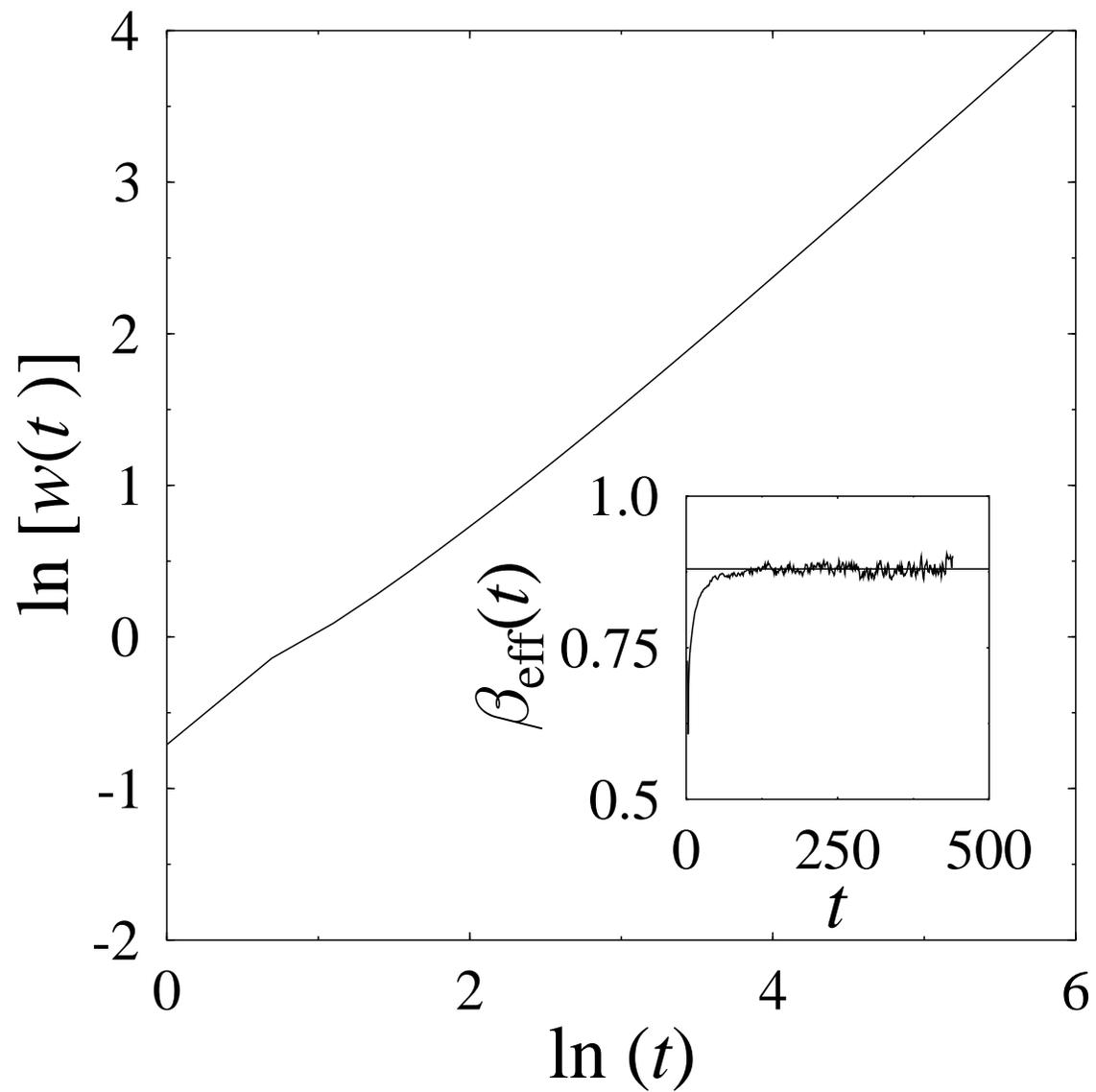
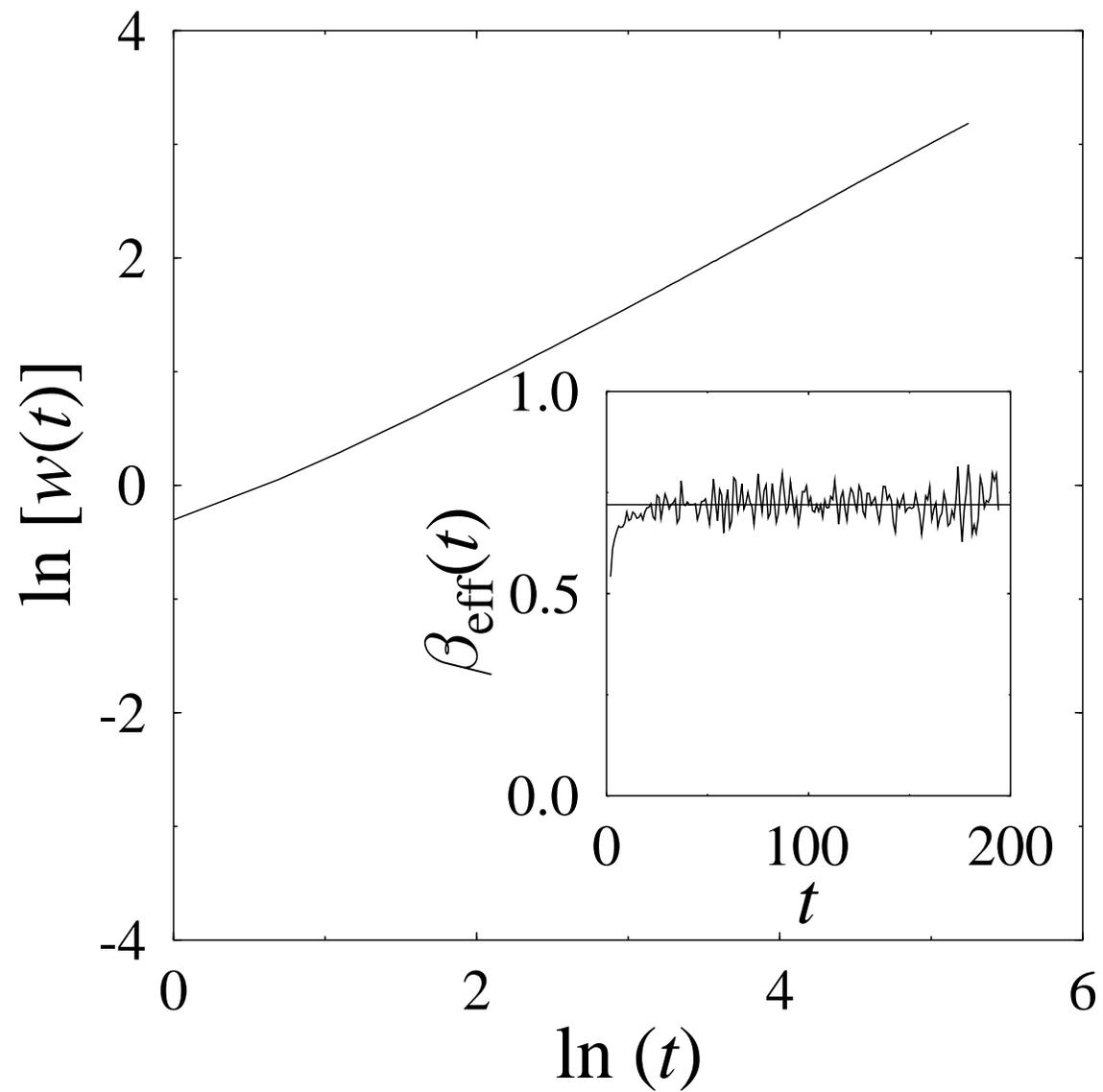
Fig. 3. (a) $w(t)$ vs. $t$ in the 2D NN lattice model
very close to the percolation transition
($c=0.59275$, $L=20000$). 
The inset shows the effective
growth exponent $\beta_{\rm eff}(t)$ and the horizontal 
line indicates the value 0.88.
(b) $w(t)$ vs. $t$ in the 3D NN lattice model very close to the
percolation transition
($c = 0.312$, $L \times L \times L = 1100 \times 1100$). The inset shows 
the effective growth exponent $\beta_{\rm eff}(t)$ vs. $t$ and the
horizontal line indicates the value 0.72. 

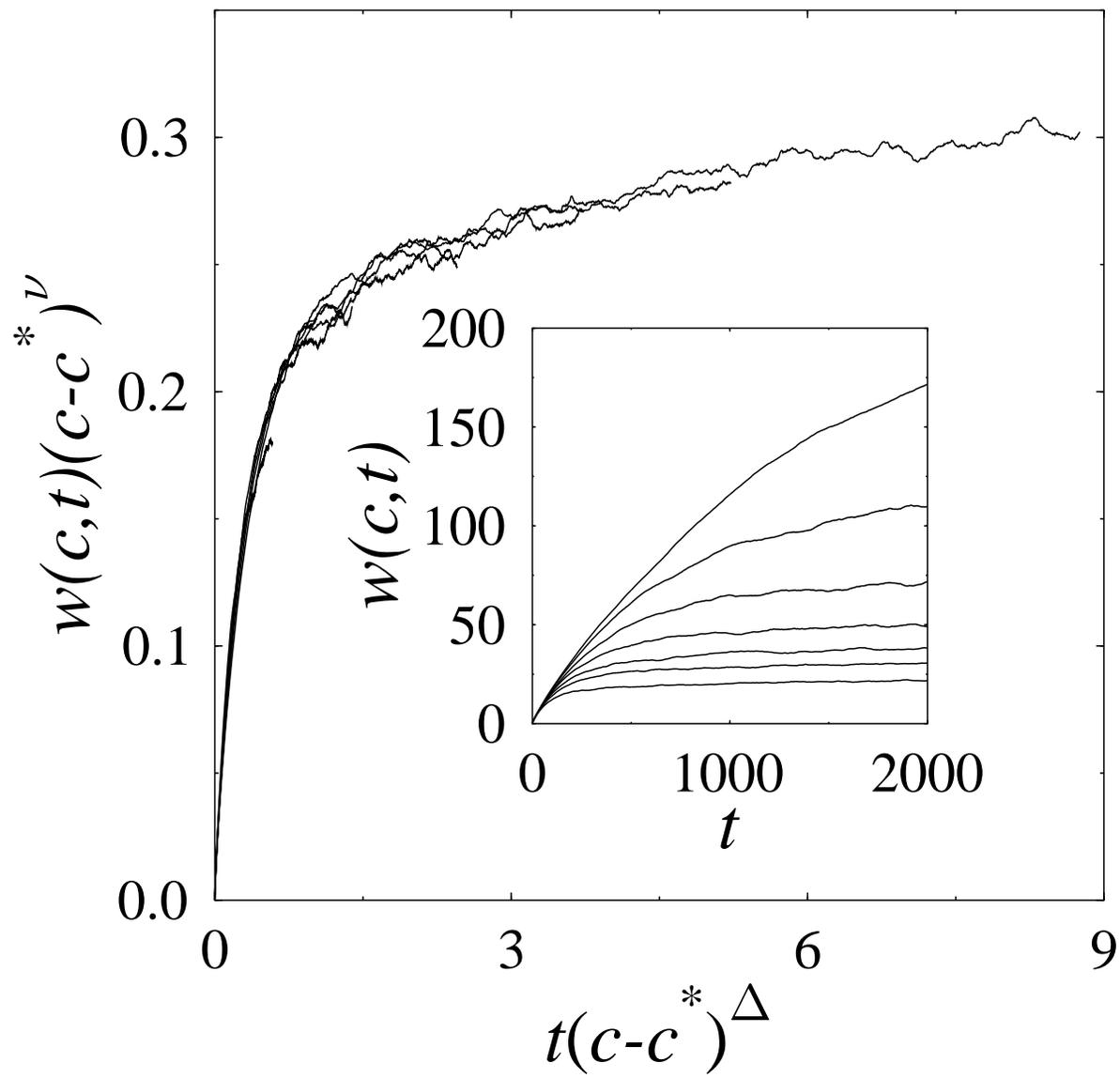
Fig. 4. Crossover scaling function $f(t/\tau_c)$
of the global width $w(c,t)$, as defined in Eq. (6).
The unscaled data for different concentrations
($c=0.594$, $0.60$, $0.605$, $0.61$, $0.615$, $0.62$, and $0.63$, 
from top to bottom, and $L=1000$) is shown in the
inset. The data collapse has been obtained using
$\nu=1.3$ and $\Delta=1.65$. See text for details.

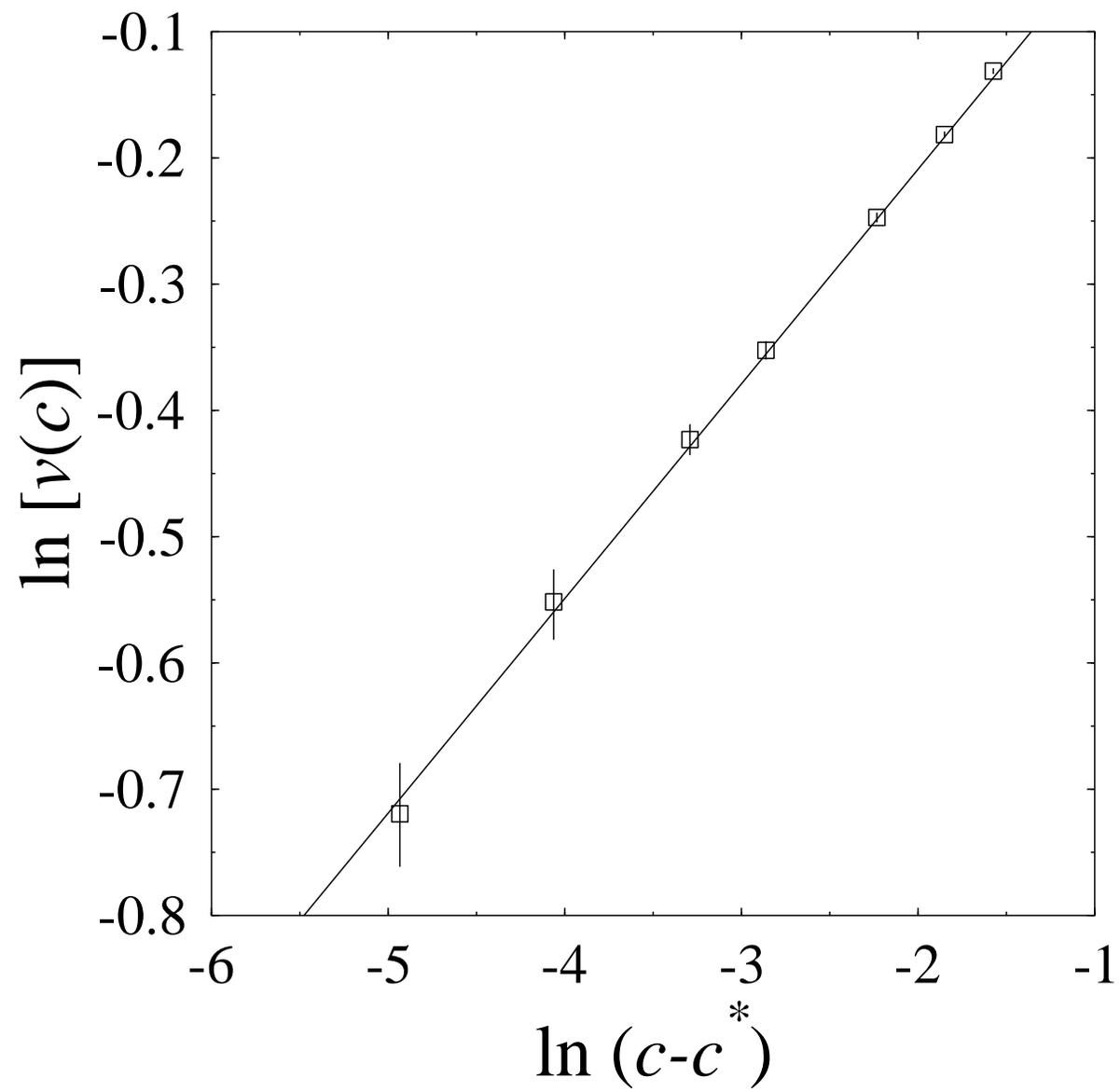
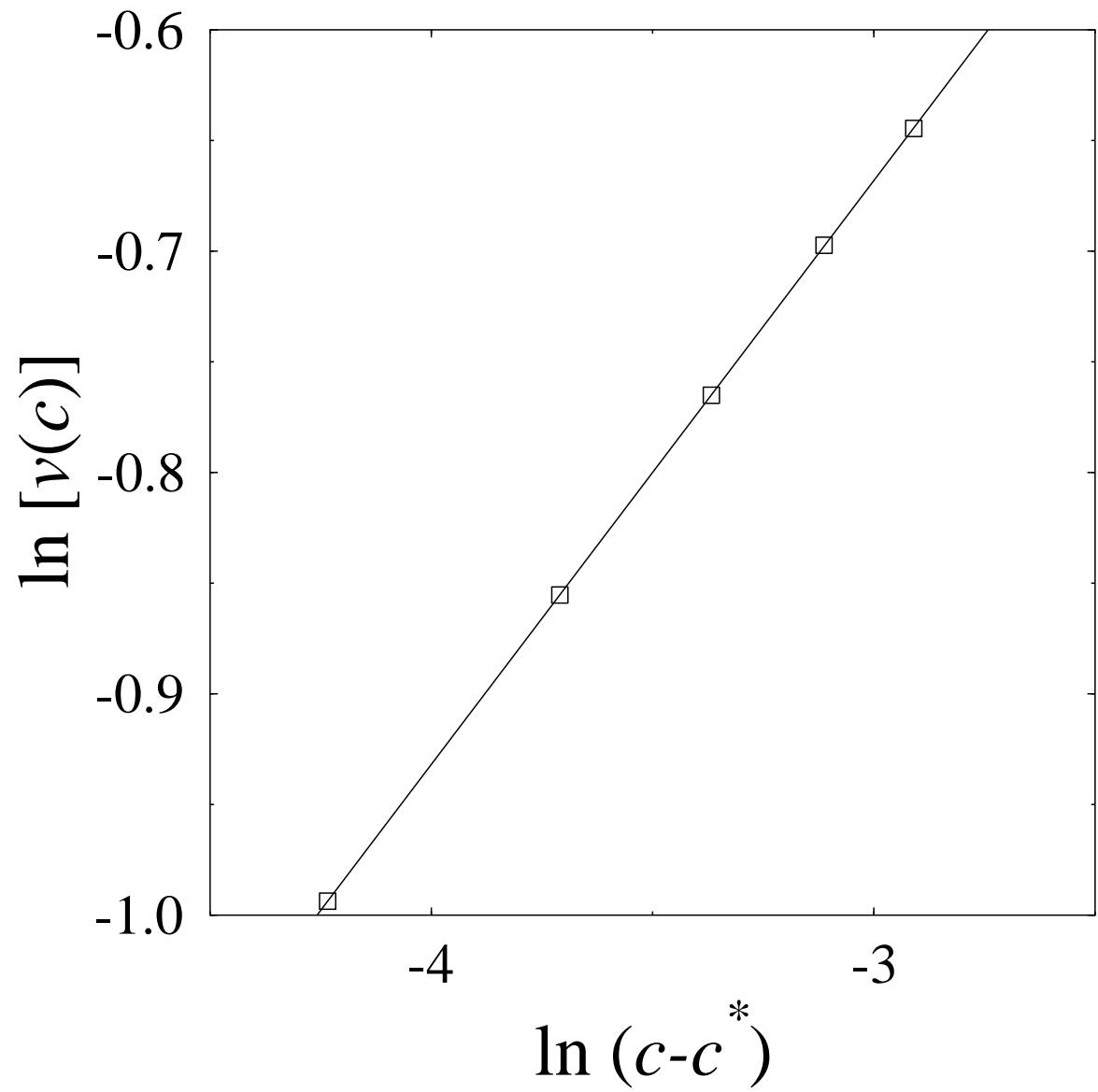
Fig. 5. (a) Scaling of the interface velocity 
$v$ vs. $c-c^*$ for the
NN 2D lattice model, with $L=2000$. The straight line shows
the best fit to the data, with $\theta=0.169$ in Eq. (9).
(b)Scaling of the interface velocity 
$v$ vs. $c-c^*$ for the
NN 3D lattice model, with $L \times L=100 \times 100$. 
The straight line shows
the best fit to the data, with $\theta=0.26$. The error
bars here are smaller than the symbol sizes. 


\begin{thebibliography}{99}
\bibitem{anotherrule}
We have also tested another definition, where the position
of the interface is given by the highest point
where {\it either} a burning tree {\it or}
a burned tree can be found. The results remain unchanged.
\bibitem{anomalous}
For some models it has been
shown that the interfaces may not be truly self-affine, and
the saturated local width
$w_{\ell} \sim \ell^{\chi_{\rm loc}}$, with
$\chi \not= \chi_{\rm loc}$ \cite{lopez97}.
Also, an anomalous temporal exponent can be defined
through the difference $\beta - \chi_{\rm loc}/z$.
However, we do not find evidence of this kind of
anomalous scaling in the present models near $c^*$,
since both the temporal and spatial scaling of the
local and global widths (and the correlation functions
for $q=2$) give the same exponents, i.e. $\beta \approx
0.88$ and $\chi \approx 0.54$.
\bibitem{huom} Naturally the system size $L$ is the third
length scale in the problem, but since we are interested
in the crossover in time in $w(c,t,L)$, we assume here
that $L$ is large enough.
\bibitem{QEW}
Incidentally, our estimates of $\beta^*$ and $\chi^*$ for the 2D
case are very close
to some of the
numerical results obtained for the Quenched Edwards-Wilkinson (QEW)
equation that is given by Eq. (1) with $\lambda=0$
\cite{amaral94,barabasi95,leschhorn96}. However,
in the QEW equation the interface is
self-affine at and close to depinning \cite{leschhorn96}, and
its scaling exponents cannot be related to percolation theory.
\end{thebibliography}
\end{document}